\documentclass[aps,showpacs,two column]{revtex4}

\usepackage{graphicx}
\usepackage[squaren]{SIunits}

\usepackage{amssymb}
\usepackage{color}
\begin{document}
\title{Structural and magnetic properties of the layered manganese oxychalcogenides (LaO)$_2$Mn$_2$Se$_2$O
and (BaF)$_2$Mn$_2$Se$_2$O}

\author{R. H. Liu, J. S. Zhang, P. Cheng, X. G. Luo, J. J. Ying, Y. J. Yan, M. Zhang, A. F. Wang, Z. J. Xiang, G. J. Ye and X. H. Chen}
\altaffiliation{Corresponding author} \email{chenxh@ustc.edu.cn}
\affiliation{Hefei  National Laboratory for Physical Science at
Microscale and Department of Physics, University of Science and
Technology of China, Hefei, Anhui 230026, People's Republic of
China}

\begin{abstract}

The new layered manganese oxychalcogenides (LaO)$_2$Mn$_2$Se$_2$O
and (BaF)$_2$Mn$_2$Se$_2$O, isostructural to (LaO)$_2$Fe$_2$Se$_2$O,
were synthesized by using solid state reaction. The single crystals
of the former compound were also successfully grown using fusion
method. The polycrytalline samples show the semiconducting behavior
with the activation energy gaps of about 278 meV and 416 meV for
(LaO)$_2$Mn$_2$Se$_2$O and (BaF)$_2$Mn$_2$Se$_2$O, respectively. The
magnetic susceptibility and specific heat indicate an
antiferromagnetic (AFM) transition at around 160$\pm$1 K for the
former compound and 100$\pm$1 K for the second compound. The strong
anisotropic magnetic properties below T$_N$ of the former compound
suggests a long-range canted AFM ordering. A broad maximum of the
susceptibility can be observed for the two compounds at high
temperatures of 360 K and 210 K, respectively, suggesting that
strong frustrated magnetic correlation gives rise to low-dimensional
AFM or short-range ordering at high temperatures in these rare
transition metal oxychalcogenides with an AFM checkerboard spin
lattice.

\end{abstract}
\vskip 15 pt \pacs{81.05$\pm$t, 75.30$\pm$m, 75.40$\pm$s,
71.27.$+$a} \maketitle


\section{Introduction}

Low-dimensional spin systems have received considerable attention
because of the possible emergence of novel physics such as
superconductivity\cite{bednorz,cava,maeno,takada}, colossal
magnetoresistance (CMR)\cite{terasaki}, giant thermoelectric power
(GTP)\cite{kimura} etc. The parent compound of high-T$_{\rm c}$
cuprates is antiferromagnetic (AFM) Mott insulator with a square
lattice of copper\cite{anderson}. The recently discovered iron-based
pnictides are the spin-density-wave type (SDW-type) AFM bad
metal\cite{kamihara,xhchen,ccruz}. They are also placed at the
boundary of itinerancy and Mott localization in many
studies\cite{qimiao1,pcdai}. Furthermore, the CMR materials also
evolve from Mott insulators by doping\cite{imada}. It is very
important to understand the correlated electron structure and
magnetic properties of those Mott insulators, which helps people to
study the novel physics. The discovery of non-copper high-T$_{\rm
c}$ iron-based pnictides with two-dimensional FeAs layers revived
intense interests in the layered transition-metal
oxychalcogenides\cite{adam} or oxypnictides\cite{mayer}. These
compounds share a common square lattice of the transition-metal. The
titanium oxypnictides $R$Ti$_2$$Pn$$_2$O ($R$ = Na$_2$, Ba,
(SrF)$_2$ and (SmO)$_2$, $Pn$ = As and
Sb)\cite{adam,tadashi1,pickett,tadashi2,tadashi3,tadashi4,lrh1,lrh2}
have a SDW/charge-density-wave (CDW) instability corresponding to an
anomalous transition in resistivity, susceptibility and Hall
coefficient etc, similar to the properties of iron pnictides. The
transition metal oxychalcogenides $R_2$Fe$_2$$Q$$_2$O ($R$ = (BaF),
(SrF) and (LaO), $Q$ = S and Se)\cite{mayer,kabbour} are
isostructural to the titanium oxypnictides. The previous studies
found that the Fe analogues formed a long-range AFM ordering below
83 K$\sim$106 K and were Mott insulators due to the narrowing of the
Fe $d$-electron bands and corresponding enhancement of correlation
effects\cite{qimiao2}. More recently, Wang et al\cite{wang}
synthesized (LaO)$_2$Co$_2$Se$_2$O, and found that it was insulator
and had an AFM transition at T$_N$ $\sim$ 220 K. Moreover, the
authors suggested that the Co analogues had an unusual low-spin(LS,
S=1/2) state of the Co$^{2+}$ ions and a corresponding orbital
polarization. However, Wu et al\cite{huawu} suggested that the
square-lattice Mott insulator La$_2$O$_2$Co$_2$Se$_2$O had more
stable high-spin(HS, S=3/2) ground state with a considerably strong
magnetic frustration.

In the present work, we synthesize other new manganese
oxychalcogenides $R_2$Mn$_2$Se$_2$O ($R$ = (LaO) and (BaF)), which
are isostructural to the previous Fe and Co analogues. In addition,
we preform the anisotropic measurements of magnetic susceptibility
in single crystal La$_2$O$_2$Mn$_2$Se$_2$O. These intrinsic
behaviors are helpful in understanding the underlying magnetism of
the transition-metal oxychalcogenides.


\section{Experimental Details}
Polycrystalline samples of (LaO)$_2$Mn$_2$Se$_2$O and
(BaF)$_2$Mn$_2$Se$_2$O were synthesized by solid state reaction
method using BaF$_2$ (3N), BaO (3N), La$_2$O$_3$ (3N), MnSe powder
as starting materials. MnSe was pre-synthesized by heating the
mixture of Se (3N) powder and Mn (3N) powder in an evacuated quartz
tube from room temperature to 750 $\celsius$ in 12 hours and staying
at 750 $\celsius$ for 24 hours. La$_2$O$_3$ was dried by heating in
air at 900 $\celsius$ for 20 hours before using. The raw materials
and precursors were weighed according to the stoichiometric ratio of
(BaF)$_2$Mn$_2$Se$_2$O or (LaO)$_2$Mn$_2$Se$_2$O, thoroughly
grounded, pressed into pellets and then sealed in evacuated quartz
tubes. The sealed tubes were sintered at 1000 $\celsius$ for 40
hours for (LaO)$_2$Mn$_2$Se$_2$O samples and 800 $\celsius$ for 60
hours for (BaF)$_2$Mn$_2$Se$_2$O samples. In order to improve their
purity and homogeneity, the resulting products were reground in
argon atmosphere before applying a second heat treatment at the same
temperature. To grow single crystal, the formed
(LaO)$_2$Mn$_2$Se$_2$O pellets were loaded into an alumina crucible
and then sealed in evacuated quartz tube. The tube was slowly heated
to 1300 $\celsius$ at a rate of 5 $\celsius$/min and kept at 1300
$\celsius$ for 10 hours. Subsequently, the temperature was lowered
slowly to 1000 $\celsius$ in 40 hours, and then the quartz tube was
cooled in the furnace by shutting off the power. The black single
crystals can be yielded at the bottom of the alumina crucible. The
sample preparation processes except for heating were carried out in
the glove box with highly pure argon atmosphere filled.

X-ray powder diffraction and single crystal diffraction were both
carried out on TTRAX3 theta/theta rotating anode X-ray
Diffractometer (Japan) with Cu K$\alpha$ radiation and a fixed
graphite monochromator. The powder diffraction data was collected in
the 5 $\sim$ 115$^{o}$ 2{\rm $\theta$} range at room temperature.
The lattice parameters and the crystal structures were refined by
using the Rietveld method in the programs GSAS package, applying
Thompson-Cox-Hastings functions with asymmetry corrections as
reflection profiles. Direct current (DC) magnetic susceptibility
measurement was performed by using a SQUID magnetometer (Quantum
Design MPMS-XL7s). The dc electrical resistivity was measured on the
Quantum Design PPMS with four probes for (LaO)$_2$Mn$_2$Se$_2$O and
two probes for (BaF)$_2$Mn$_2$Se$_2$O.


\section{Results and Discussion}

\begin{figure}[h]
\includegraphics[width=0.48\textwidth]{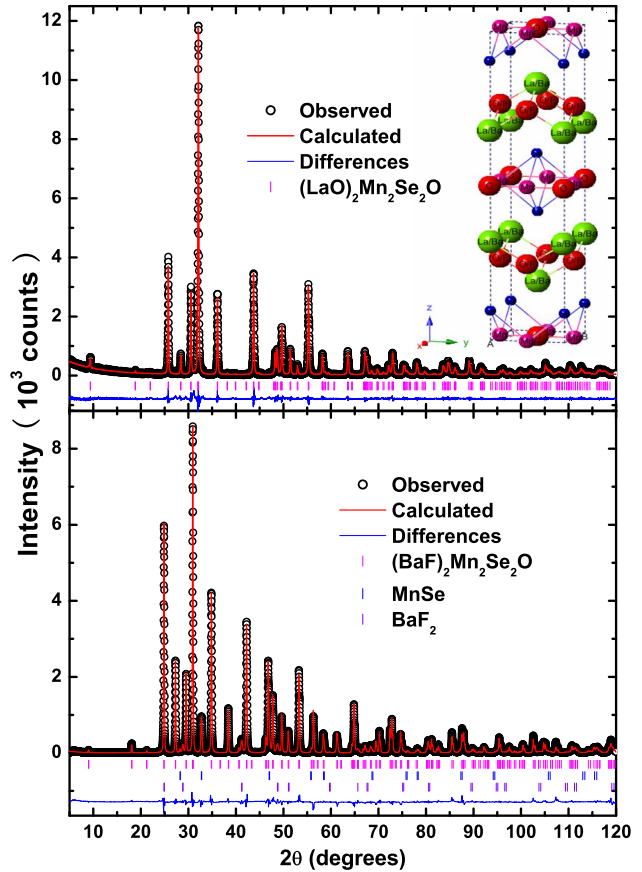}
\caption{X-ray powder diffraction patterns at room temperature for (LaO)$_2$Mn$_2$Se$_2$O
 (top panel), and (BaF)$_2$Mn$_2$Se$_2$O (bottom panel). The solid line indicates the intensities
  calculated using the Rietveld method. The bottom solid curves show the differences between the
   observed and calculated intensities. The vertical short lines indicate the Bragg peak positions
    of the target compounds, respectively. Inset of top panel: Crystal structure of $R_2$Mn$_2$Se$_2$O ($R$ = (LaO) and (BaF)).}
\end{figure}


\begin{table}[h]
\caption{ Atomic coordinates and equivalent isotropic displacement
parameters (U$_{iso}$) for $R_2$Mn$_2$Se$_2$O ($R$ = (LaO) and
(BaF)).}
\begin{center}
\tabcolsep 0pt \vspace*{-12pt}
\def\temptablewidth{0.48\textwidth}
{\rule{\temptablewidth}{1pt}}
\begin{tabular*}{\temptablewidth}{@{\extracolsep{\fill}}cccccc}
    atom& site& x  & y & z& U$_{iso}$({\AA}$^2$)\\
    \hline
    & & (LaO)$_2$Mn$_2$Se$_2$O &  &    \\
       La& 4e& 1/2 & 1/2  & 0.18626(4) &  0.0017(3)  \\
       O & 4d& 1/2 & 0    & 1/4        &  0.004(2)  \\
       Mn& 4c& 1/2 & 0    & 0          &  0.0121(7)  \\
       Se& 4e& 0   & 0    & 0.09984(7) &  0.0076(4)  \\
       O & 2a& 1/2 &1/2   &0           &  0.033(5)  \\

       &&   (BaF)$_2$Mn$_2$Se$_2$O &&     \\
       Ba & 4e&1/2 & 1/2  & 0.17207(5)&0.0209(1) \\
       F & 4d &1/2 & 0   & 1/4       & 0.0437(2)  \\
       Mn & 4c&1/2 & 0   & 0         & 0.0246(5)  \\
       Se & 4e& 0& 0 &   0.09445(4)   &0.0277(2)   \\
       O & 2a&1/2&1/2&0&               0.0148(4)    \\

       \end{tabular*}
       {\rule{\temptablewidth}{1pt}}
       \end{center}
        \end{table}

\begin{table}[h]
\caption{ Cell parameters , selected interatomic distances, bond
angles and reliability factors obtained using Rietveld refinements
of $R_2$Mn$_2$Se$_2$O ($R$ = (LaO) and (BaF)) (Space group I4/mmm).
}
\begin{center}
 \tabcolsep 0pt \vspace*{-12pt}
\def\temptablewidth{0.48\textwidth}
{\rule{\temptablewidth}{1pt}}
\begin{tabular*}{\temptablewidth}{@{\extracolsep{\fill}}ccccc}
    & (LaO)$_2$Mn$_2$Se$_2$O & (BaF)$_2$Mn$_2$Se$_2$O  \\
    \hline
       Space group&I4/mmm & I4/mmm   \\
       a({\AA})& 4.14097(7)& 4.2756(1)    \\
       c({\AA})& 18.8588(4)& 19.5919(4)    \\
       V({\AA}$^3$) & 323.38(2)&  358.73(2)  \\
       Z  & 2& 2   \\
       Data points &11500&11500\\
       R$_{wp}$(\%)& 0.1067 & 0.1150     \\
       R$_{p}$(\%) & 0.0751 & 0.1066     \\

       Bond lengths({\AA})&&      \\
       d$_{La-O}$ or d$_{Ba-F}$   &2.3941(4)$\times$4& 2.62701(3)$\times$4  \\
       d$_{La-Se}$ or d$_{Ba-Se}$  &3.3511(7)$\times$4& 3.38429(4)$\times$4   \\
       d$_{Mn-O}$  &2.07049(4)$\times$2 & 2.13784(3) $\times$2  \\
       d$_{Mn-Se}$ &2.7986(9)$\times$4 & 2.82752(3)$\times$4     \\
       d$_{Mn-Mn}$ &2.92811(5)$\times$4& 3.02336(5)$\times$4     \\

       Bond angles({$^o$})&&      \\
       O-La-O or F-Ba-F    &75.399(1)$\times$4& 70.261(1)$\times$4  \\
                           &119.724(1)$\times$4& 108.936(1)$\times$4  \\
       Mn-Se-Mn            &63.086(1)$\times$4& 64.639(1)$\times$4  \\
                           &95.434(1)$\times$4& 98.242(1)$\times$4  \\
       \end{tabular*}
       {\rule{\temptablewidth}{1pt}}
       \end{center}
        \end{table}

\begin{figure}[h]
\includegraphics[width=0.48\textwidth]{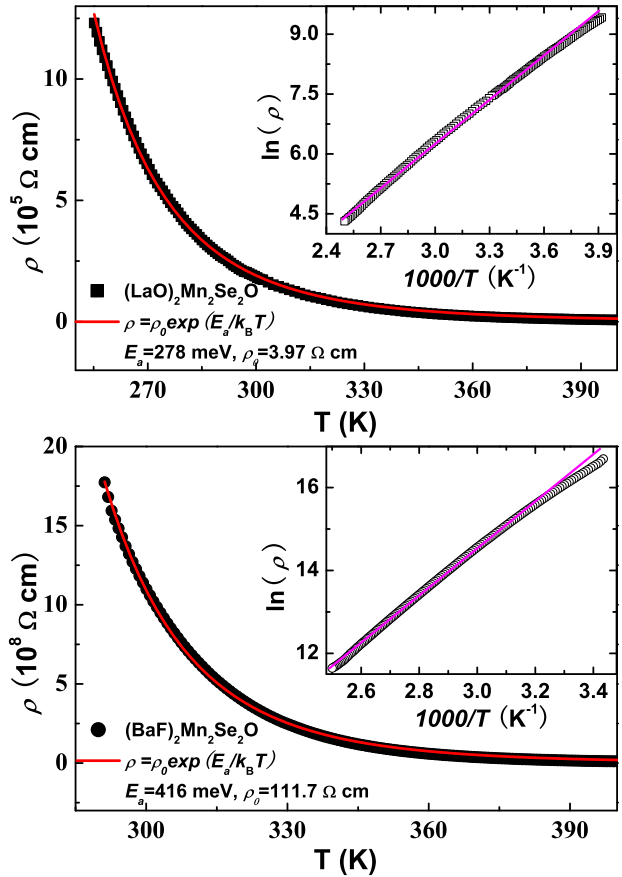}
\caption{Temperature dependence of resistivity \emph{$\rho$} for
polycrystalline (LaO)$_2$Mn$_2$Se$_2$O (top panel) and
(BaF)$_2$Mn$_2$Se$_2$O (bottom panel). The solid line indicates
the curves fitted using the Arrhenius equation
\emph{$\rho$=$\rho$$_0$exp(E$_a$/k$_{\rm B}$T)}. Insets: The natural
logarithm of resistivity \emph{{\rm ln} $\rho$} plotted against reciprocal of temperature.}
\end{figure}

\begin{figure}[t]
\includegraphics[width=0.48\textwidth]{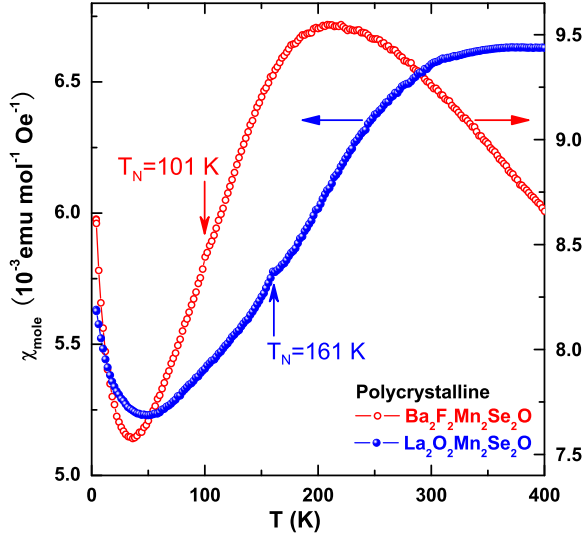}
\caption{Temperature dependence of magnetic susceptibility measured under
H= 5 T for polycrystalline samples of $R$$_2$Mn$_2$Se$_2$O ($R$ = (LaO) and (BaF)).}
\end{figure}

\begin{figure}[t]
\includegraphics[width=0.48\textwidth]{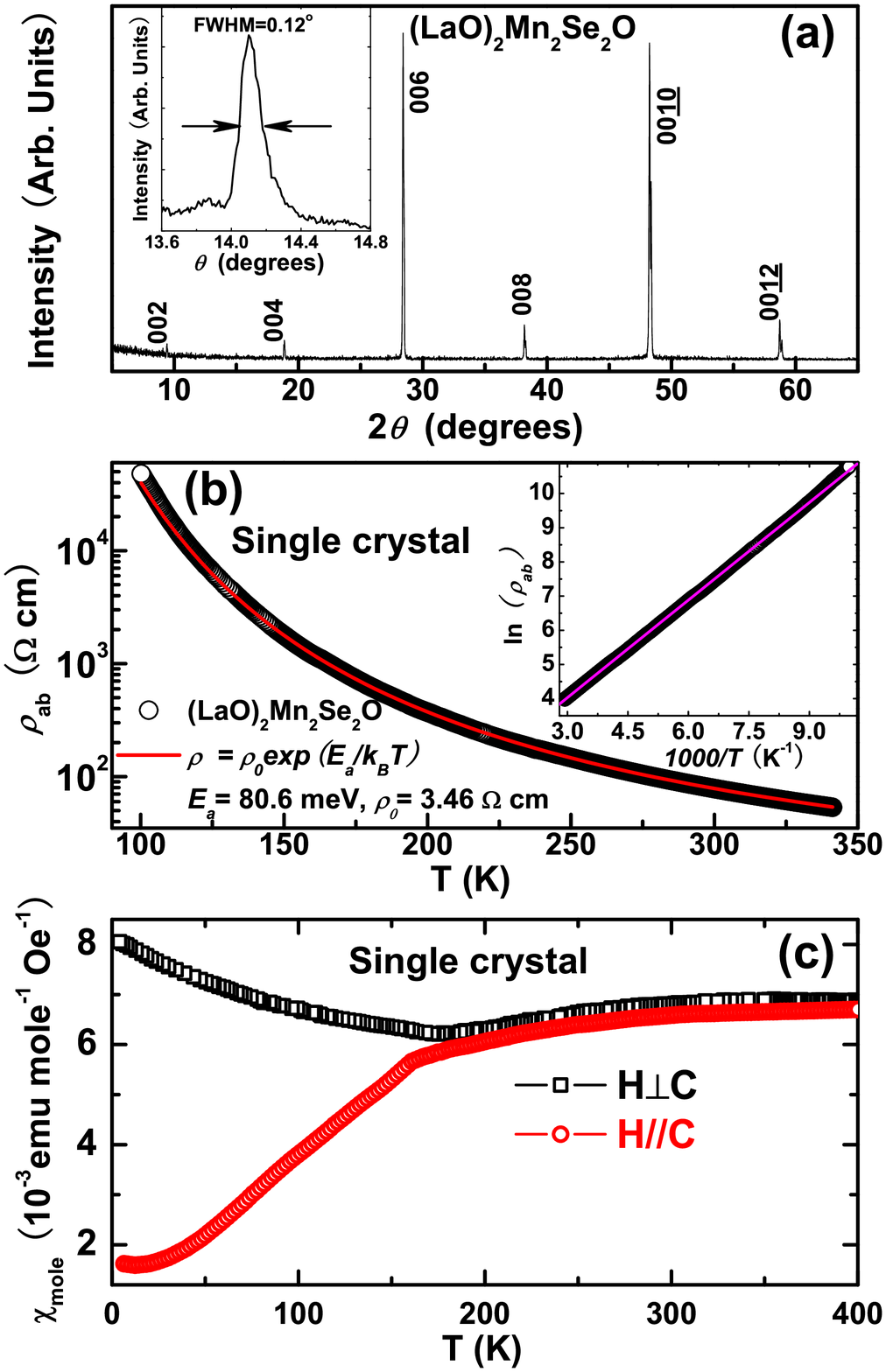}
\caption{(a) Single crystal x-ray diffraction pattern of (LaO)$_2$Mn$_2$Se$_2$O. Only (00$l$)
 diffraction peaks show up, suggesting that the c axis is perpendicular to the plane of the plate;
  Inset of top panel: rocking curve of (006) reflection. (b) Temperature dependence of resistivity \emph{$\rho_{ab}$} for
single crystal (LaO)$_2$Mn$_2$Se$_2$O. Insets: The natural logarithm
of resistivity \emph{{\rm ln} $\rho$} plotted against reciprocal of
temperature. (c) Temperature dependence of magnetic
  susceptibility measured under H= 7 T applied along the \emph{ab}-plane(black), along the
  \emph{c}-axis(red) of (LaO)$_2$Mn$_2$Se$_2$O single crystal, respectively. }
\end{figure}

\subsection{Structure Refinements}
Figure 1 shows X-ray powder diffraction patterns for the samples
$R_2$Mn$_2$Se$_2$O ($R$ = (LaO) (top panel) and (BaF) (bottom
panel)). All the XRD reflections for the former compound,
(LaO)$_2$Mn$_2$Se$_2$O, can be indexed with a tetragonal unit cell.
The data were well fitted by using the structure model of
(LaO)$_2$Fe$_2$Se$_2$O with the space group of \emph{I4/mmm}. As
shown in inset of Fig.1, those Mn-based compounds are isostructural
with Fe and Co oxychalcogenides and Ti oxypnictides. These layered
compounds are stacked alternately with the fluorite type
[$Ln$$_2$O$_2$] or [$AE$$_2$F$_2$] ($Ln$= rare earth metals, and
$AE$=alkali earth metals) and edge-shared $M$$_2$$Pn$$_2$O layers
($M$ = transition metal, $Pn$= chalcogenide and pnictide). In the
$M$$_2$$Pn$$_2$O unit, $M$$_2$O forms a square planar with
anticonfiguration to the CuO$_2$ layer of high-T$_{\rm c}$ cuprates,
while two $Pn$ are located above and below the center of this square
unit. For such Se/O mixed tetragonal structure, according to
Goodenough-Kanamori rules\cite{goodenough1,goodenough2}, there are
three intralayer spin exchange interactions: AF excharge $J_1$ via
corner sharing Mn-O-Mn, AF exchange $J_2$ via face sharing
Mn-O-Mn/Mn-Se-Mn and FM exchange $J_3$ via edge sharing Mn-Se-Mn.
These spin exchange interactions induce magnetic frustration in
these compounds with checkerboard spin lattice. According to our
Rietveld refinement, we also know that, for (BaF)$_2$Mn$_2$Se$_2$O,
there are little impurity phases of BaF$_2$ and MnSe in XRD
patterns. The data were well fitted by the targeted tetragonal phase
and these impurity phases in structure refinement. The refined
lattice parameters ($a$ = 4.14097(7)${\rm \AA}$ and $c$ =
18.8588(4)${\rm \AA}$ for (LaO)$_2$Mn$_2$Se$_2$O; $a$ =
4.2756(1)${\rm \AA}$ and $c$ = 19.5919(4)${\rm \AA}$ for
(BaF)$_2$Mn$_2$Se$_2$O) are larger than those of the Fe and Co
analogues due to Mn$^{2+}$ ions being relatively larger than
Fe$^{2+}$ and Co$^{2+}$ ions. The interatomic distance of Mn-Mn
(2.928 ${\rm \AA})$ is smaller than that of rock salt structure MnO
(3.142 ${\rm \AA}$) with an T$_{\rm N}$$\sim$118 K antiferromagnetic
order\cite{shull}. The detailed results of the Rietveld refinements
are listed in Tables 1 and 2.

To get more information by comparing with other analogues, we
measured temperature dependence of resistivity for the as-prepared
(LaO)$_2$Mn$_2$Se$_2$O and (BaF)$_2$Mn$_2$Se$_2$O samples and found
that they both show semiconducting behavior. The room temperature
resistivity is as high as $\sim$ 10$^5$ and 10$^9$ $\Omega$ cm,
respectively, being three or six orders of magnitude higher than
those of (LaO)$_2$Fe$_2$Se$_2$O (10$^2$ $\Omega$ cm) and
(BaF)$_2$Fe$_2$Se$_2$O (10$^3$ $\Omega$ cm). In addition, the
resistivity obeys thermally activated behavior(as shown in inset of
Fig.2), and can be fitted by the Arrhenius equation:
\emph{$\rho$=$\rho$$_0$exp(E$_a$/k$_B$T)}, where $\rho$$_0$ is the
pre-exponential factor and \emph{k}$_B$ is the Boltzmann constant.
The obtained activation energy $E_a$ is 278 meV for
(LaO)$_2$Mn$_2$Se$_2$O, which is located between Fe analogue (190
meV) and Co anologue (350 meV). The previous studies indicated that
the Fe and Co analogous were Mott insulators with narrowing
3$\emph{d}$ electronic bands due to strong correlation effects. Our
results may suggest that the Mn oxychalcogenides are also
correlation-induced insulators with low-energy spin excitations
(AFM) at low temperatures. For (BaF)$_2$Mn$_2$Se$_2$O, there is
larger activation energy (416 meV) with higher resistivity. It can
be explained from the viewpoint of electron-hopping between adjacent
spin sites. From the Rietveld refinement results above, one can
easily see that the (LaO)$_2$Mn$_2$Se$_2$O has a greater integral
for hopping between adjacent Mn$^{2+}$ sites because it has shorter
Mn...Mn, Mn-Se-Mn and Mn-O-Mn distances than those of
(BaF)$_2$Mn$_2$Se$_2$O.

Figure 3 shows the temperature dependence of the DC magnetic
susceptibility $\chi(T)$ obtained at 5 T for (LaO)$_2$Mn$_2$Se$_2$O
and (BaF)$_2$Mn$_2$Se$_2$O polycrystalline samples with the
temperature ranging from 4 K to 400 K. The $\chi(T)$ of
polycrystalline (LaO)$_2$Mn$_2$Se$_2$O shows a broad maximum around
360 K, suggesting the existence of a low-dimensional
antiferromagnetism or a two-dimensional short-range ordering. Below
360 K, $\chi(T)$ decreases with reducing temperature to 50 K. In
this decreasing $\chi(T)$, a small anomaly is observed at 161 K,
which may suggest a magnetic transition occurring at this
temperature. A broad maximum can also be observed in the $\chi(T)$
for polycrystalline (BaF)$_2$Mn$_2$Se$_2$O sample, at a temperature
of 210 K. The different temperature corresponding to the maximum
$\chi(T)$ for the two compounds indicate that (LaO)$_2$Mn$_2$Se$_2$O
has a stronger AF coupling than (BaF)$_2$Mn$_2$Se$_2$O does.
Combining with structural parameters listed in Table 2, it suggests
that the chemical pressure effect enhances the spin exchange
interactions. A similar behavior has been observed in isostructural
Fe analogues\cite{kabbour}. A very slight anomaly also is observed
around 101 K for $\chi(T)$ of (BaF)$_2$Mn$_2$Se$_2$O. However, the
obvious Curie paramagnetic tail at low temperature may arise from
the small amount of impurity phase.

To shed light on the nature of the anomaly at 161 K in $\chi(T)$ of
(LaO)$_2$Mn$_2$Se$_2$O, we measured the DC magnetic susceptibility
on the single crystal (LaO)$_2$Mn$_2$Se$_2$O sample with the
magnetic field applied in $ab$-plane and along $c$-axis. From the
XRD patterns in Fig. 4a, only (00$l$) reflections are observed for
the single crystal, suggesting that the plate-like single crystal is
grown along $c$-axis. The full width of half maximum (FWHM) in the
rocking curve of the (006) reflection is  0.12 $^\circ$, suggesting
the high quality of the single crystal. Elemental analysis obtained
by Energy Dispersive X-ray Spectroscopy (EDX) gives the atomic ratio
of La : O : Mn : Se is roughly 24.21 : 33.98 : 22.02 : 19.79,
suggesting the single crystal has a little Se vacancies due to high
temperature melt growing. Figure 4b shows the temperature dependence
of resistivity \emph{$\rho_{ab}$} for the single crystal sample. It
also show semiconducting behavior as same as polycrystalline sample.
The room temperature resistivity of single crystal
(\emph{$\rho_{ab}$}=10$^2$ $\Omega$ cm) is three orders of magnitude
smaller than that of polycrystalline sample. The obtained activation
energy \textit{E$_a$} is 80.6 meV. There are several factors that
lead to a huge difference of resistivity between single crystal and
polycrystal. Firstly, there are a little Se vacancies in the single
crystal sample, which will lead some carriers into system and
increase its conductivity. Secondly, the resistance of
polycrystalline sample is an average of out-plane and in-plane
resistance. However, the in-plane conductance should be larger than
that of out-plane since these compounds have a two-dimensional
layered structure. In addition, it is more difficult to obtain an
accurate resistivity for polycrystalline sample due to the impact of
grain boundary effect and loose density of pressed powder sample.
The susceptibility of the single crystal is shown at Fig. 4c. The
high-temperature $\chi(T)$ shows a broad maximum around 350 K when
magnetic field is applied in $ab$-plane, which is a little lower
than that observed in the polycrystalline sample, and monotonically
decreases as the field is applied along $c$-axis. With further
decreasing temperature, a pronounced downward kink is observed at
161 K in $\chi(T)$ when field is applied along \emph{c}-axis,
indicating an AFM transition with $T_{\rm N}$ =  161 K. However, an
obvious increase appears below this temperature in $\chi(T)$ as
field applied within $ab$-plane. This anisotropic behavior can be
understood within a scenario of a canted AFM ordering, that is, the
spins ordered along c-axis antiferromagnetically, while canted in
the $ab$-plane. Therefore, the in-plane component of spins is
aligned ferromagnetically with the magnetic field applied within
$ab$-plane. Several magnetic structure models (AFM1 or AFM6 with
spin align c-axis) have been suggested in the Fe analogue and Co
analogue by first principle
calculation\cite{kabbour,qimiao2,wang,huawu}. However, Free et
al.\cite{david} suggested AFM3-type with majority spin direction in
ab plane magnetic structure in (LaO)$_2$Fe$_2$Se$_2$O, similar to
that in FeTe\cite{weibao}. To completely understand the magnetic
structure, people need detailed neutron scattering experiment to
confirm for Mn analogues.

\begin{figure}[t]
\includegraphics[width=0.48\textwidth]{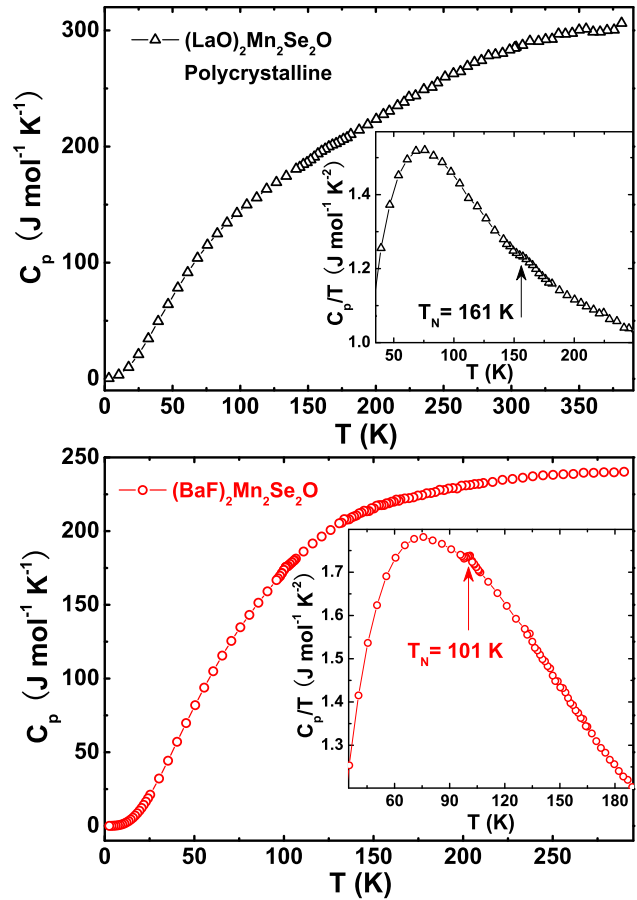}
\caption{Temperature dependence of specific heat for polycrystalline
(LaO)$_2$Mn$_2$Se$_2$O(top panel) and (BaF)$_2$Mn$_2$Se$_2$O (bottom
panel). There are weak anomaly around 161 K for
(LaO)$_2$Mn$_2$Se$_2$O and 101 K for (BaF)$_2$Mn$_2$Se$_2$O in
$C_p/T$ vs. T curves, shown in the insets of top and bottom panel,
respectively.}
\end{figure}

In order to further confirm the magnetic transition, the heat
capacity is measured between 2 K and 400 K by a relaxation-time
method in PPMS. There is a very slight anomaly around 161 K for
(LaO)$_2$Mn$_2$Se$_2$O and 101 K for (BaF)$_2$Mn$_2$Se$_2$O in
specific heat measurement, respectively, shown in Fig. 5, being
consistent with the susceptibility anomaly. To easily distinguish
specific heat anomaly, the $C_p/T$ vs. $T$ curves are plotted in the
inset of Fig.5. These magnetic and calorimetric anomalies suggest
the existence of a three dimensional long range order AFM transition
evolved from high temperature low dimensional AFM or short range
ordering. Comparing with Mn analogues, the Fe analogues and Co
analogues had a sharp peak in specific heat around $T_N$. It
suggests that there is more strong frustrated magnetic correlation
in Mn analogues than Fe or Co analogues above $T_N$. It maybe leads
to a low dimension AFM order in Mn analogues above $T_N$ so that
only a partial magnetic entropy is lost below the anomalous
temperature.

\section{Conclusion}
The new layered manganese oxychalcogenides (LaO)$_2$Mn$_2$Se$_2$O
and (BaF)$_2$Mn$_2$Se$_2$O were successfully synthesized. Their
crystal structures were refined by using the model isostructural to
(LaO)$_2$Fe$_2$Se$_2$O. We also grew the single crystals of the
former compound using high temperature flux method. The
susceptibility of as-grown single crystal samples have strong
anisotropic magnetic properties below $T_{\rm N}$, and the spins
spontaneously align predominantly along the $\emph{c}$-axis. While
being an analogue to the isostructural iron and cobalt
oxychalcogenides, Mn-based compounds maybe also belong to a Mott
insulator with an antiferromagnetic ground state. A broad maximum of
the susceptibility observed above $T_{\rm N}$
 suggests that these two-dimensional checkerboard spin lattice with a
 considerably strong magnetic frustration maybe form a short-range
    magnetic ordering or low-dimensional AFM ordering above
 $T_{\rm N}$. These intrinsic magnetic behaviors are helpful in understanding the underlying physics
 of these rare transition metal oxychalcogenides or oxypnictides
 with a frustrated AFM checkerboard spin lattice.

\section{Acknowledgment}

This work is supported by the National Natural Science Foundation of
China(973 Program No:2011CB00101 and Grant No. 51021091), the
Ministry of Science and Technology of China, and Chinese Academy of
Sciences.
\\

{\bf Notes:} As we are preparing the manuscript, we note that the
compound (LaO)$_2$Mn$_2$Se$_2$O has been synthesized by Ni {\sl et
al.} and a G-type AFM order has been suggested by neutron scattering
experiment at low temperature, in accepted articles list of Physical
Review B.

\end{document}